\documentclass[aps,prd,twocolumn,groupedaddress]{revtex4}
\usepackage{epsfig,dcolumn}
\usepackage{graphicx}
\usepackage{bm}
\newcommand{\nc}{\newcommand}
\nc{\mb}[1]{\makebox[#1]{}}
\nc{\CC}{{\scriptscriptstyle CC}}
\nc{\NC}{{\scriptscriptstyle NC}}
\nc{\V}{{\rm v}}
\nc{\SC}{{\rm s}}
\nc{\W}{{\scriptscriptstyle W}}
\nc{\X}{{\scriptscriptstyle X}}
\nc{\Z}{{\scriptscriptstyle Z}}
\nc{\F}{{\scriptscriptstyle F}}
\nc{\C}{{\scriptscriptstyle C}}
\nc{\Scr}{{\scriptscriptstyle S}}
\nc{\CS}{{\scriptscriptstyle CS}}
\nc{\SV}{{\scriptscriptstyle SV}}
\nc{\DY}{{\scriptscriptstyle DY}}
\nc{\PW}{{\scriptscriptstyle PW}}
\nc{\SB}{{\scriptscriptstyle SB}}
\nc{\CSV}{{\scriptscriptstyle CSV}}
\nc{\QCD}{{\scriptscriptstyle QCD}}
\nc{\GLS}{{\scriptscriptstyle GLS}}
\nc{\CIB}{{\scriptscriptstyle CIB}}
\nc{\PT}{{\scriptscriptstyle PT}}
\nc{\ASYM}{{\scriptscriptstyle asym}}
\nc{\IE}{{\it i.e.,\ }}
\nc{\EG}{{\it e.g.,\ }}
\nc{\EA}{{\it et al.}}
\nc{\AH}{{\it ad hoc\ }}
\nc{\CHPT}{{$\chi_{\PT}$\ }}

\nc{\NCA}{{\em Nuovo Cimento}}
\nc{\NIM}{{\em Nucl. Instrum. Methods}}
\nc{\NIMA}{{\em Nucl. Instrum. Methods} A}
\nc{\NPB}{{\em Nucl. Phys.} B}
\nc{\PLB}{{\em Phys. Lett.}  B}
\nc{\PRL}{{\em Phys. Rev. Lett.}}

\nc{\PRD}{{\em Phys. Rev.} D}
\nc{\PRC}{{\em Phys. Rev.} C}
\nc{\ZPC}{{\em Z. Phys.} C}

\nc{\st}{\scriptstyle}
\nc{\sst}{\scriptscriptstyle}
\nc{\mco}{\multicolumn}
\nc{\epp}{\epsilon^{\prime}}
\nc{\vep}{\varepsilon}
\nc{\ra}{\rightarrow}
\nc{\ppg}{\pi^+\pi^-\gamma}
\nc{\xpi}{x_{\pi}}
\nc{\xF}{x_{\F}}
\nc{\pis}{\pi_{\SC}(\xpi)}
\nc{\piv}{\pi_{\V}(\xpi)}
\nc{\pist}{\widetilde{\pi}_{\SC}(\xpi)}
\nc{\nuN}{{\nu N_0}}
\nc{\nubN}{{\overline{\nu} N_0}}
\nc{\ovnu}{{\overline{\nu}}}
\nc{\dvx}{{d_{\V}(x)}}
\nc{\deldv}{{\delta \dvx}}
\nc{\uvx}{{u_{\V}(x)}}
\nc{\qvx}{{q_{\V}(x)}}
\nc{\deluv}{{\delta \uvx}}
\nc{\bux}{{\bar{u}(x)}}
\nc{\bdx}{{\bar{d}(x)}}
\nc{\dub}{{\delta \bux}}
\nc{\ddb}{{\delta \bdx}}
\nc{\buo}{{\bar{u}(x_1)}}
\nc{\bdo}{{\bar{d}(x_1)}}
\nc{\dubo}{{\delta \buo}}
\nc{\ddbo}{{\delta \bdo}}
\nc{\but}{{\bar{u}(x_2)}}
\nc{\bdt}{{\bar{d}(x_2)}}
\nc{\dubt}{{\delta \but}}
\nc{\ddbt}{{\delta \bdt}}
\nc{\deluo}{{\delta u(x_1)}}
\nc{\deldo}{{\delta d(x_1)}}
\nc{\delut}{{\delta u(x_2)}}
\nc{\deldt}{{\delta d(x_2)}}
\nc{\snuNC}{{\langle \sigma^{\nuN}_{\NC}\rangle }}
\nc{\snubNC}{{\langle \sigma^{\nubN}_{\NC}\rangle }}
\nc{\snuCC}{{\langle \sigma^{\nuN}_{\CC}\rangle }}
\nc{\snubCC}{{\langle \sigma^{\nubN}_{\CC}\rangle }}
\nc{\Rnu}{{R^{\nu}}}
\nc{\Rnub}{{R^{\overline{\nu}}}}
\nc{\sintW}{{\sin^2 \theta_{\W} }}
\nc{\costC}{{\cos^2 \Theta_{\C} }}
\nc{\sintC}{{\sin^2 \Theta_{\C} }}
\nc{\MS}{{\overline{MS}}}
\nc{\vp}{{\bf p}}
\nc{\rz}{{\rho_0^2}}
\nc{\ko}{K^0}
\nc{\kb}{\bar{K^0}}
\nc{\al}{\alpha}
\nc{\ab}{\bar{\alpha}}
\nc{\be}{\begin{equation}}
\nc{\ee}{\end{equation}}
\nc{\bea}{\begin{eqnarray}}
\nc{\eea}{\end{eqnarray}}
\nc{\beast}{\begin{eqnarray*}}
\nc{\eeast}{\end{eqnarray*}}

\begin{document}

\titlepage

\title{Parton Charge Symmetry Violation: Electromagnetic Effects 
and W Production Asymmetries} 

\author{J.T.Londergan}

\email{tlonderg@indiana.edu}
\affiliation{Department of Physics and Nuclear
            Theory Center,\\ Indiana University,\\ 
            Bloomington, IN 47405, USA}

\author{D.P. Murdock}
\email{murdock@tntech.edu}
\affiliation {Department of Physics,\\ Tennessee Technological University\\ 
                Cookeville, TN 38505, USA}

\author{A.W.Thomas}
\email{awthomas@jlab.org}
\affiliation {Jefferson Lab, 12000 Jefferson Ave.,\\ 
                Newport News, VA 23606, USA}
\date{\today}


\begin{abstract} 
{}Recent phenomenological work has examined two different ways of including 
charge symmetry violation in  parton distribution functions. First, a global 
phenomenological 
fit to high energy data has included charge symmetry breaking terms, leading 
to limits on the magnitude of parton charge symmetry breaking. In a second 
approach, two groups have included the coupling of partons to photons 
in the QCD evolution equations. One possible experiment that could search for 
isospin violation in parton distributions is a measurement of the 
asymmetry in W production at 
a collider. In this work we include both of the postulated 
sources of parton charge symmetry 
violation. We show that, given charge symmetry violation 
of a magnitude consistent 
with existing high energy data, the expected W production asymmetries would 
be quite small, generally less than one percent. 
\end{abstract}




\maketitle



Charge symmetry represents a specific form of isospin 
invariance (a rotation of $180^\circ$ about the ``2'' axis 
in isospin space) that is quite well respected at low 
energies~\cite{Miller,Henley}. Since there is no direct experimental evidence 
of charge symmetry violation (CSV) for PDFs \cite{Lon98,Lon04}, it was 
reasonable, at least in the beginning, 
to assume that it held as well for parton distribution functions 
(PDFs). However, we know that small violations of charge symmetry do  
arise from both the mass differences of light current quarks, and from 
electromagnetic effects. There have been some theoretical estimates of 
charge symmetry violation in PDFs, and recently charge 
symmetry violation has been included in phenomenological PDFs. 
{}Furthermore, the estimated size of the CSV is such that it can produce 
important effects in some experiments, for example in precise tests of 
physics beyond the Standard Model~\cite{Londergan:2005xt,Mil05}.

Global fits of PDFs by Martin, Roberts, Stirling and Thorne 
(MRST)~\cite{MRST03} included the possibility 
of charge symmetry violating PDFs for valence and sea quarks.  By 
construction, the resulting parton distributions will 
agree with the array of experimental data used in global fits. 
The valence quark CSV PDFs were chosen to have the specific form   
\bea 
 \delta u_{\V}(x) &=& - \delta d_{\V}(x) = \kappa (1-x)^4 x^{-0.5}\, 
  (x - .0909)  \nonumber \\ 
  \delta u_{\V}(x) &=& u_{\V}^p(x) - d_{\V}^n(x) \ ; \hspace{0.25cm} 
  \delta d_{\V}(x) = d_{\V}^p(x) - u_{\V}^n(x) \nonumber \\ 
\label{eq:CSVmrst}
\eea
At both small and large $x$ the valence quark CSV term is qualitatively 
similar to phenomenological valence quark distributions \cite{MRST01}, and 
the first moment of the valence CSV distribution is zero, a necessary 
condition to preserve valence quark 
normalization. The single coefficient $\kappa$ was varied in the global fit 
to high energy data. To minimize the resulting computing time, MRST neglected 
the $Q^2$ dependence of this CSV effect.  The best value they obtained was 
$\kappa = -0.2$, and the 90\% confidence level included the range 
$-0.8 \le \kappa \le +0.65$. It is interesting to note that for the best 
fit obtained by MRST, the valence 
quark CSV distributions are in very good agreement both in sign and 
magnitude with predictions from quark model 
calculations~\cite{Sat92,Rod94}  -- see also the model independent 
constraint on the second moment obtained 
in Refs.~\cite{Londergan:2003ij,Londergan:2003pq}.   

In a separate global fit to the same data, MRST included the possibility of 
sea quark CSV effects. The MRST functional form chosen for sea quark CSV was 
\bea 
\bar{u}^n(x) &=& \bar{d}^p(x)\left[ 1 + \delta \right] \nonumber \\  
\bar{d}^n(x) &=& \bar{u}^p(x)\left[ 1 - \delta \right] 
\label{eq:seaCSV}
\eea  
This form was chosen to insure that the total momentum carried by antiquarks 
in the neutron and proton was approximately equal. Once again, they assumed no 
$Q^2$ dependence for these CSV distributions.The best fit was obtained 
for $\delta = 0.08$.  

An alternative, phenomenological approach to the problem of charge symmetry 
violation associated with the electromagnetic interaction has been 
proposed by both MRST \cite{MRST05} and Glueck, Jimenez-Delgado
and Reya \cite{Glu05}. By analogy with the usual QCD evolution involving 
gluon radiation, these authors suggested that one assume charge symmetry 
at some initial low-mass scale, and include in the evolution 
equations the effect of photon radiation.
When one includes QED contributions in this way, to lowest order in both the 
strong coupling $\alpha_{\Scr}$ and the EM coupling $\alpha$, the so-called 
DGLAP evolution equations due to Dokshitzer 
\cite{Dok77}, Gribov and Lipatov \cite{Gri72} and Altarelli and 
Parisi \cite{Alt77} are modified. The MRST group obtains  
\bea
{\partial q_i (x,\mu^2)\over \partial {\rm log} \,\mu^2} &=& 
 {\alpha_{\Scr}\over 2\pi} \left[ P_{qq}\otimes q_i + P_{qg}\otimes g 
 \right] \nonumber \\ &+& {\alpha\over 2\pi} \left[ 
  \widetilde{P}_{qq}\otimes e_i^2 q_i + 
  P_{q\gamma}\otimes e_i^2 \gamma \right] \ , \nonumber \\ 
{\partial g (x,\mu^2)\over \partial {\rm log} \,\mu^2} &=& 
 {\alpha_{\Scr}\over 2\pi} \left[ P_{gq}\otimes \sum_j q_j + P_{gg}\otimes g 
 \right] \ , \nonumber \\ 
{\partial \gamma (x,\mu^2)\over \partial {\rm log} \,\mu^2} &=& 
 {\alpha\over 2\pi} \left[ P_{\gamma q}\otimes \sum_j e_j^2 q_j 
  + P_{\gamma\gamma}\otimes \gamma \right] \ . \nonumber \\ 
\label{eq:QEDevol}
\eea
In Eq.~\ref{eq:QEDevol}, the right hand side of the schematic evolution 
equations represents a convolution of the splitting functions with the quark 
and gluon distributions (which have an explicit dependence on the 
factorization scale parameter $\mu^2$).  Inclusion of the electromagnetic 
contribution 
to QCD evolution introduces a ``photon parton 
distribution'' $\gamma(x,\mu^2)$ which is coupled to the quark and 
gluon distributions. The new splitting functions that occur in 
Eq.~\ref{eq:QEDevol} are related to the standard QCD splitting functions by 
\bea
\widetilde{P}_{qq} &=& P_{qq}/C_F; \hspace{1.0cm} P_{\gamma q} = P_{gq}/C_F 
\nonumber \\ P_{q\gamma} &=& P_{qg}/T_R; \hspace{1.0cm} P_{\gamma \gamma} = 
   -{2\over 3}\,\sum_i e_i^2 \delta(1-y) 
\label{eq:split}
\eea
Conservation of momentum is assured by the relation 
\be
\int_0^1\, dx \, x \, \left[ \sum_i q_i(x,\mu^2) + g(x,\mu^2) + 
   \gamma(x, \mu^2) \right] = 1 
\label{eq:momcons}
\ee

It is necessary to simplify Eqs.~\ref{eq:QEDevol}.  First, since the 
EM interaction is not asymptotically free, it is not clear how to 
set the starting values for the various PDFs that are coupled by 
these QED effects. In particular, it is not clear where the QED effects 
should be assumed to vanish. Second, inclusion of the QED couplings could in 
principle more than double the number of parton 
distribution functions (one must 
now differentiate between proton and neutron PDFs, in addition to the 
new photon parton distributions).  Two groups have adopted somewhat 
different strategies, with similar overall results. 
Glueck \EA~\cite{Glu05} point out that the photon parton distribution is 
already of order $\alpha$, as is clear by inspection of 
Eq.~\ref{eq:QEDevol}.  Consequently to leading order in $\alpha$ they    
drop terms involving $\gamma(x,\mu^2)$ from the right-hand side of 
Eq.~\ref{eq:QEDevol}.  When they adopt the standard convention for DIS 
reactions of setting the scale $\mu^2 = Q^2$, Glueck \EA~then obtain 
convolution equations for the charge symmetry violating valence quark 
distributions arising from QED coupling, 
\bea
&\,&{d \over d\, {\rm ln}Q^2}\,\delta u_{\V}(x,Q^2) = {\alpha\over 2\pi} 
  \,\int_x^1 {dy\over y}\, P\left({x\over y}\right) u_{\V}(y,Q^2) 
   \nonumber \\ &\,&{d \over d\, {\rm ln}Q^2}\,\delta d_{\V}(x,Q^2) = 
  -{\alpha\over 2\pi} \,\int_x^1 {dy\over y}\, P\left({x\over y}\right) 
  d_{\V}(y,Q^2) \nonumber \\ 
  &\,&P(z) = (e_u^2 - e_d^2)\, P_{qq}(z) = (e_u^2 - e_d^2) \left({1+z^2 
  \over 1-z} \right)_+
\label{eq:Gluevol}
\eea
Similar relations hold for the antiquark distributions. Glueck \EA~assume 
that the average current quark mass $\overline{m}_q$, taken as 10 MeV, is 
the kinematical lower bound for a photon emitted by a quark. This is 
analogous to taking the electron mass as the lower limit for radiation 
of photons in the earliest calculations of the Lamb shift (before the 
advent of renormalization group arguments) \cite{Bet57}.      
Eq.~\ref{eq:Gluevol} is then integrated from $\overline{m}_q^2$ to $Q^2$.  
The rationale here is to evaluate QED evolution effects while keeping 
the QCD effects fixed.  Thus, the quark distributions appearing on the right 
hand side of Eq.~\ref{eq:Gluevol} are taken from the GRV leading-order parton 
distributions \cite{GRV98}. In the resulting integrals, in the region 
$q^2 < \mu_{LO}^2= 0.26$ GeV$^2$ corresponding to momentum transfers below the 
input scale for GRV, the PDFs are ``frozen,'' \IE in this region they are 
assumed to be equal to their value at the input scale $\mu_{LO}^2$.  

\begin{figure}
\includegraphics[width=3.3in]{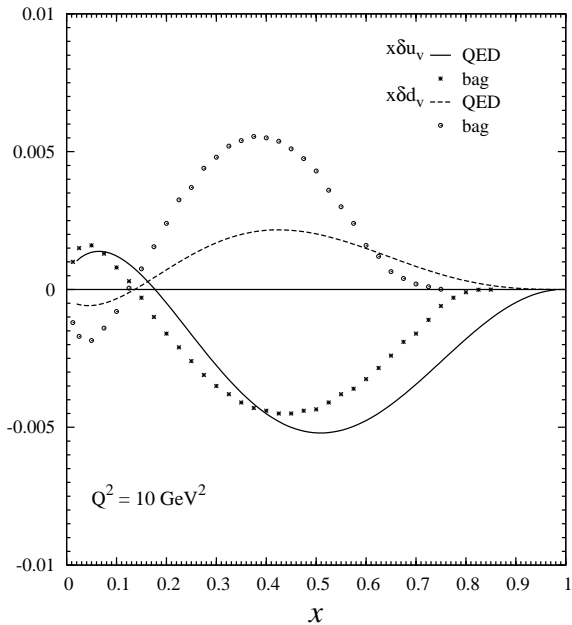}
\caption{The isospin-violating majority $x\delta u_{\V}$ (solid curve) 
and minority $x\delta d_{\V}$ (dashed curve) valence parton distributions 
obtained by Glueck \EA \ \protect{\cite{Glu05}} at $Q^2 = 10$ GeV$^2$, 
assuming QED evolution from a scale set by the current quark mass. 
These are compared with majority (solid points) 
and minority (open circles) CSV distributions obtained from theoretical 
bag model calculations \protect{\cite{Rod94}}. 
\label{Fig:Glueck}}
\end{figure}

The resulting valence isospin asymmetries $x\delta u_{\V}$ and 
$x\delta d_{\V}$ are plotted in Fig.~\ref{Fig:Glueck} at $Q^2 = 10$ GeV$^2$.  
For comparison, they are plotted along with the valence quark isospin 
asymmetries obtained by Rodionov \EA~\cite{Rod94,Lon03}.  The latter CSV 
distributions were obtained from bag model calculations, where  
charge symmetry violation was assumed to arise from mass 
differences of the residual diquarks $\delta \widetilde{m} = m_{dd} - m_{uu}$ 
and from the target nucleon mass difference $\delta M = M_n - M_p$. The 
quantity $\delta \widetilde{m}$ was taken as 4 MeV \cite{Sat92,Rod94}, which 
includes an estimate of the EM contribution to this mass difference. While 
the quantity $\delta u_{\V}$ is quite similar in both sign and magnitude for 
both the bag model and the QED calculations, the QED results for 
$\delta d_{\V}$ are roughly half as large as the bag model results. As noted 
previously, the bag model results for valence quark CSV are extremely close to 
those obtained by MRST using the phenomenological form of Eq.~\ref{eq:CSVmrst},
for the best-fit value $\kappa = -0.2$.     

The MRST group \cite{MRST05} solves the evolution equations of 
Eq.~\ref{eq:QEDevol} with assumptions about the parton distributions 
at the starting scale $Q_0^2 = 1$ GeV$^2$. At the starting scale, the sea 
quark and gluon distributions are assumed to be isospin symmetric. The 
starting photon parton distributions are taken as those due to one-photon 
radiation from valence quarks in leading-logarithm approximation, evolved from 
current quark masses $m_u = 6$ MeV and $m_d = 10$ MeV  to $Q_0$. This 
produces different photon PDFs for neutron and proton at the starting scale.  
Enforcing overall quark momentum conservation from Eq.~\ref{eq:momcons} 
requires valence quark isospin asymmetry at the starting scale.  MRST assume
that this takes the form
\be 
 d_{\V}^n - u_{\V}^p = 2(d_{\V}^p - u_{\V}^n) = \epsilon 
  (u_{\V}^p - 2d_{\V}^p) \ \ .
\label{eq:epsdef}
\ee
Eq.~\ref{eq:epsdef} is a simple phenomenological form chosen to obey the 
valence quark normalization condition, and it produces isospin violating 
distributions that resemble the valence PDFs at large and small $x$.  The 
parameter $\epsilon$ is determined from the overall quark momentum 
conservation condition. 

Having determined the starting distributions for the photon parton 
distribution, and the asymmetry parameter $\epsilon$, the proton's 
quark and gluon distributions at the starting scale $Q_0^2$ are 
determined in the same way as for other MRST global fits.  The only 
change is that separate DGLAP evolution equations are used for partons 
in the neutron and proton. 

If we adopt the MRST functional form for charge symmetry violating PDFs, we 
can estimate the magnitude of effects one might expect in a dedicated 
experiment designed to test parton charge symmetry. In a recent paper
\cite{Lon05d}, we estimated the magnitude of effects in two promising 
experiments. The first was a comparison of Drell-Yan cross sections 
induced by charged pions on an isoscalar target (\EG the deuteron).  The 
second experiment involved semi-inclusive deep inelastic 
scattering involving charged pion production in $e-D$ interactions. 

In this report, we consider another possible experimental test of parton 
charge symmetry. This involves measurements of $W$ production at hadron 
colliders, specifically $W$-boson production in high energy $p-D$ 
collisions. This was initially suggested by Vigdor \cite{Vig97}.  
Boros \EA~made estimates of the effects that might be expected at 
colliders such as RHIC and LHC \cite{Bor99}, and concluded that one might 
expect several percent effects in certain observables. However, these 
effects occurred because the authors had assumed very large charge 
symmetry violation in the parton sea. This large sea quark CSV was 
necessary to account for significant discrepancies between the $F_2$ 
structure functions 
extracted from high energy $\mu-D$ interactions measured by the NMC 
Collaboration \cite{Ama91,Arn97}, and the $F_2$ from $\nu-Fe$ DIS 
measured by CCFR \cite{Sel97}. However, these discrepancies disappeared 
when the neutrino reactions were re-analyzed~\cite{Boros:1999fy,Yan01}.  

There were 
two reasons for significant changes of $F_2^{\nu}$ upon re-analysis. 
First, experimental neutrino cross-sections measure a combination of 
$F_2$ and $xF_3$.  In the initial analysis the quantity $xF_3$ was calculated 
from phenomenological PDFs. In the re-analysis, this quantity was extracted 
from experiment, by using the fact that the two structure functions have 
different $y$ dependences in the cross sections. The value of $xF_3$ that was 
extracted differed 
considerably from the phenomenological $xF_3$, and this subsequently 
changed the value of $F_2$ that was extracted. The second significant 
change arose through the use of next-to-leading order (NLO) equations 
for charm quark mass effects~\cite{Boros:1999fy}, 
rather than the ``slow re-scaling'' model 
\cite{Bar76,Geo76}. Although the 
MRST analysis found evidence for sea quark CSV \cite{MRST03}, it was 
considerably smaller than that extracted from the original data, including 
the slow re-scaling contribution. 

The cross sections for the processes $p+D \rightarrow W^+ + X$ and 
$p+D \rightarrow W^- + X$ have the form 
\bea 
&\,&{d\,\sigma \over d\xF}(pD \rightarrow W^+ X) \sim \costC [ 
  u(x_1)( \but + \bdt \nonumber \\ &\,& - \dubt ) + \bdo \left( u(x_2) 
  + d(x_2) - \deldt \right) ] + \nonumber \\ &\,& \sintC \left[ 
 2u(x_1)s(x_2) + s(x_1) \left( u(x_2) + d(x_2) - \deldt \right) \right] 
  \nonumber \\ 
&\,&{d\,\sigma \over d\xF}(pD \rightarrow W^- X) \sim \costC [ 
  \buo ( u(x_2) + d(x_2) \nonumber \\ &\,& - \delut ) + d(x_1) \left( 
  \but + \bdt - \ddbt \right) ] + \nonumber \\ &\,& \sintC \left[ 
  2\buo s(x_2) + s(x_1) \left( \but + \bdt - \ddbt \right) \right] 
  \nonumber \\  
\label{eq:dsigdxf}
\eea

In the absence of CSV terms, if we take the sum of the $W^+$ and $W^-$ cross 
sections, 
\be 
 \sigma{\Scr}(\xF) \equiv \left({d\,\sigma \over d\xF}\right)^{W^+} + 
   \left({d\,\sigma \over d\xF}\right)^{W^-} \ \ ,
\label{eq:sigmaSS} 
\ee
then the Cabibbo favored terms in $\sigma_{\Scr}$ are invariant under the 
exchange $x_1 \leftrightarrow x_2$, or alternatively under the transformation 
$\xF \rightarrow -\xF$, where $\xF = x_1 - x_2$. Consequently, we define the 
forward-backward asymmetry $A(\xF)$ as 
\be 
A(\xF) = { \sigma_{\Scr}(\xF) - \sigma_{\Scr}(-\xF) \over \sigma_{\Scr}(\xF) + 
 \sigma_{\Scr}(-\xF) }
\label{eq:AxF}
\ee
The only terms remaining in the quantity $A(\xF)$ are charge symmetry 
violating terms, plus terms containing strange quarks in the Cabibbo-unfavored 
sector. 

We have calculated the effects to be expected for the forward-backward 
asymmetry $A(\xF)$ for $W$-production at a hadron collider, using the charge 
symmetry violating PDFs calculated by the MRST group. We have used  
PDFs corresponding to three different sources of charge symmetry violation.  
First, we used the valence quark and sea quark CSV PDFs extracted by 
the MRST group from global fits to high energy data \cite{MRST03}.  Then 
we have added the CSV PDFs calculated by the MRST group by including the 
``QED'' contributions to QCD evolution, with assumptions about the 
parton distributions at the starting scale $Q_0^2 = 1$ GeV$^2$ 
\cite{MRST05,DurWeb}. Note that the various CSV PDFs were extracted using 
different procedures.  The 
valence and sea quark PDFs were calculated in independent global fits to 
high energy data.  In these fits, one assumed a particular functional form 
for the valence (or sea) quark CSV PDFs, which depended upon an overall 
variable parameter for the strength.  That parameter was determined by 
minimizing the $\chi^2$ of the global fit. For simplicity MRST 
neglected the $Q^2$ dependence of the CSV distributions. 

In these various global fits, one obtains different valence parton 
distributions in the minimization process (the sea quark and gluon 
distributions were essentially identical to those obtained assuming charge 
symmetry).  So the best fit valence quark PDFs obtained by MRST when they 
allowed valence quark CSV differed somewhat from those obtained when they 
allowed sea quark CSV.  In addition, the MRST global fits that allowed parton 
CSV did not explicitly include the ``QED'' contributions to QCD evolution.  
We explore the various contributions to the $W$ 
production asymmetry, despite some questions regarding the consistency in 
the different parton distributions that give rise to CSV effects.      

\begin{figure}
\includegraphics[width=3.3in]{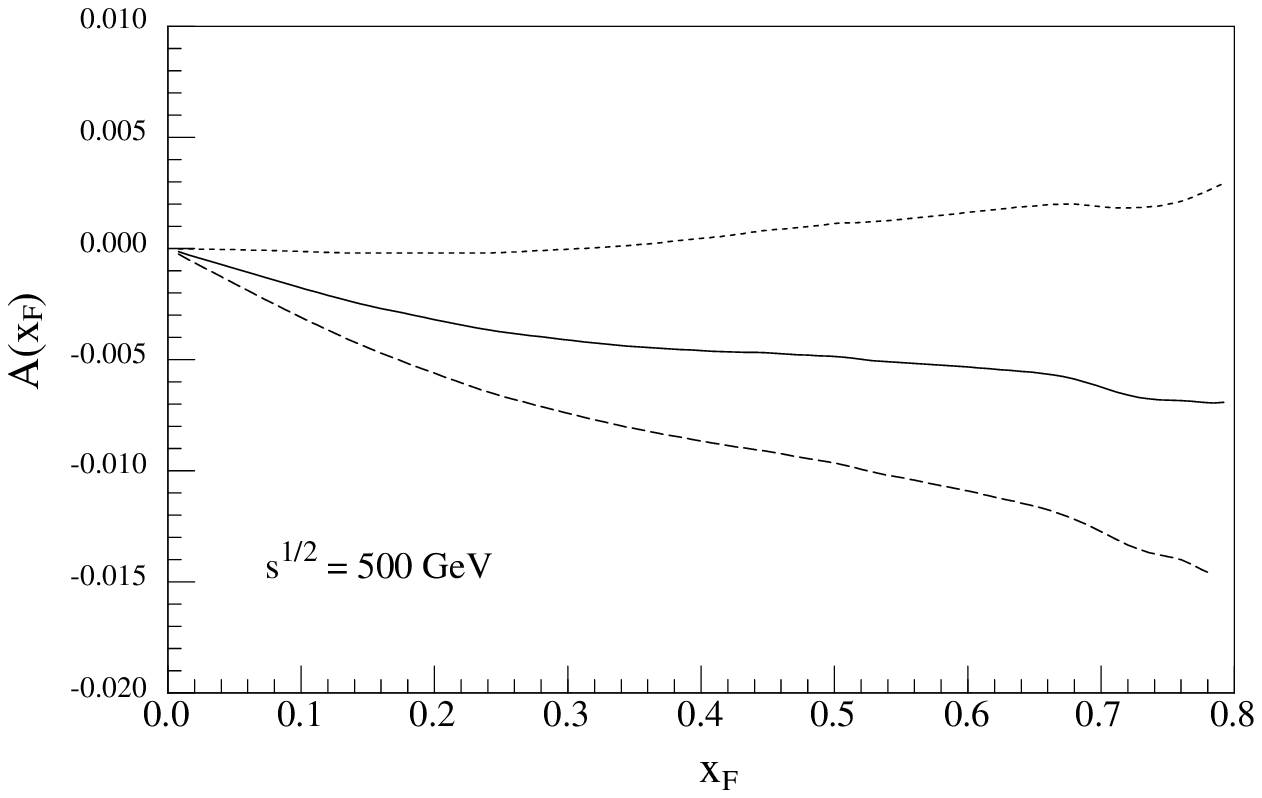}
\includegraphics[width=3.3in]{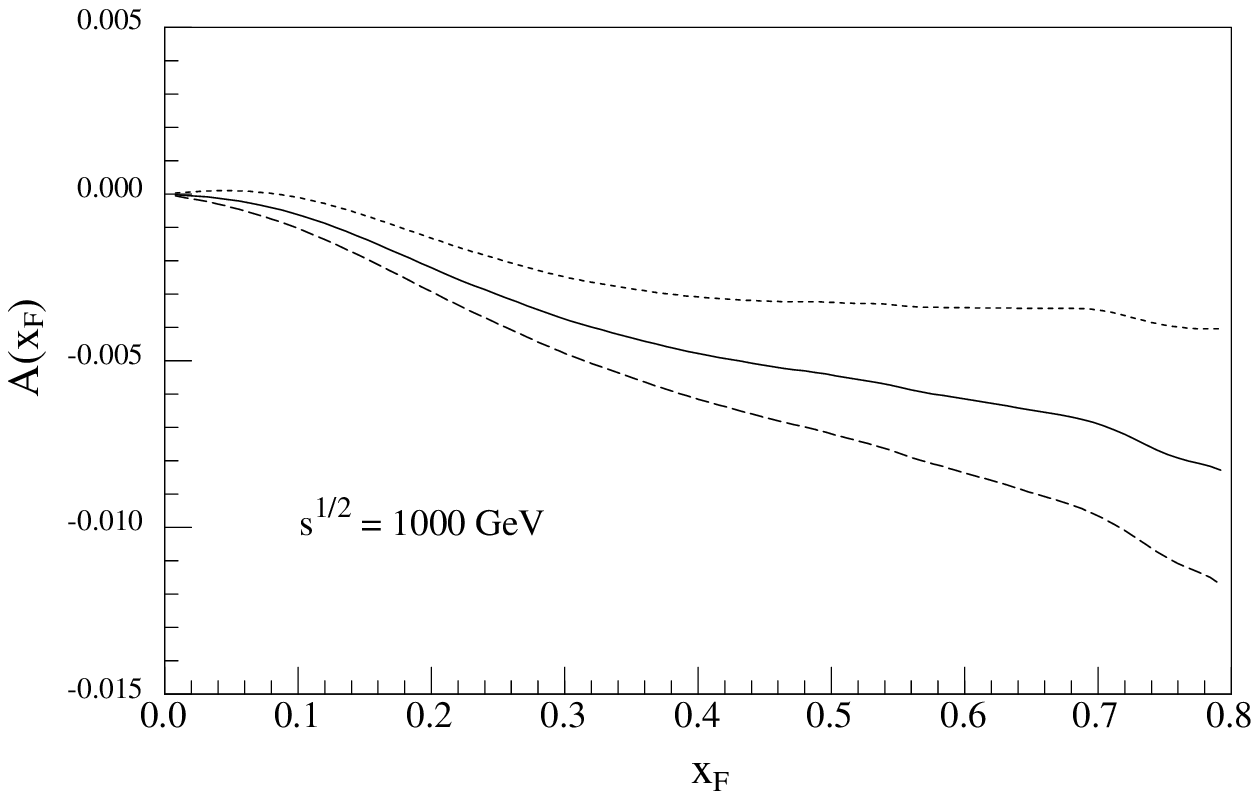}
\caption{The forward-backward asymmetry $A(\xF)$ defined in 
Eq.~(\protect\ref{eq:AxF}) as a function of $\xF$. Top graph: $\sqrt{s} = 
500$ GeV; bottom graph: $\sqrt{s} = 1000$ GeV. The curves include CSV 
generated by QED effects, and sea quark CSV described by 
Eq.~\protect{\ref{eq:seaCSV}}. They differ in the amount of valence quark 
CSV defined by Eq.~\protect{\ref{eq:CSVmrst}}. Solid curve: no valence 
quark CSV, $\kappa = 0$; long-dashed curve: $\kappa = +0.65$; short-dashed 
curve: $\kappa = -0.8$.
\label{Fig:Afig}}
\end{figure}

In Fig.~\ref{Fig:Afig}, we plot the forward-backward asymmetry expected 
for $W$ production, as defined in Eq.~\ref{eq:AxF}.  The parton 
distribution functions are obtained from the MRST analysis that includes 
the ``QED'' contribution to DGLAP evolution; this analysis includes 
electromagnetic 
couplings in the evolution equations that give rise to isospin violation 
\cite{MRST05}, as given by Eq.~\ref{eq:QEDevol}.  The input data for 
these global fits was that used in the MRST2004 analysis
\cite{MRST04}. The top figure is calculated for $\sqrt{s} = 500$ GeV, and the 
bottom figure is calculated for $\sqrt{s} = 1000$ GeV. All three curves in 
Fig.~\ref{Fig:Afig} include the sea quark 
CSV terms from Eq.~\ref{eq:seaCSV}.  They differ in the amount of valence 
quark CSV of the type given by Eq.~\ref{eq:CSVmrst}.  In each graph, the 
solid line corresponds to no valence CSV ($\kappa = 0$), while the 
short-dashed and long-dashed curves 
are for $\kappa = -0.8$ and $\kappa = +0.65$, the parameters  
corresponding to the 90\% confidence limit obtained in the MRST global fit. 

Fig.~\ref{Fig:Afig} shows that the sea quark CSV and the QED 
isospin violation tend to produce negative forward-backward $W$ asymmetries.  
The contributions from valence quark CSV, with magnitudes at the 90\% 
confidence level extracted by MRST, contribute a roughly equal magnitude 
to the asymmetries produced by the other sources of CSV. For negative values 
of $\kappa$, which agree with theoretical estimates of valence quark CSV 
\cite{Sat92,Rod94}, the valence CSV terms tend 
to cancel the asymmetry produced by sea quark CSV; while for positive 
values of $\kappa$ the various sources of isospin violation tend to add.    
Note that the predicted forward-backward asymmetries are quite small. The 
magnitude 
is less than $0.01$ for almost all values of $\xF$.  These results are 
considerably smaller than those obtained by Boros \EA~\cite{Bor99}, who 
predicted rather large positive values for $A(\xF)$, as large as 
$A(\xF) \sim +0.07$ for $\xF \sim 0.7$. There are two reasons for this 
difference.  First, the sea quark CSV terms are substantially smaller 
for MRST than those extracted by Boros \EA, by a factor of five or six; the 
sea quark CSV terms obtained by MRST and Boros also have opposite signs.  
Second, with the very large values of sea quark CSV extracted by Boros, the 
Cabibbo unfavored terms were negligible. However, with the much smaller sea 
quark CSV obtained by MRST, the Cabibbo unfavored terms can no longer be 
neglected, and they tend to cancel the Cabibbo favored contribution. 
 
This calculation of the forward-backward asymmetry for $W$ production finds 
that the expected effects are quite small, generally 
less than one percent for the range of CSV effects obtained by MRST. With 
the expected levels of isospin violation in PDFs, it will be necessary to 
measure these asymmetries to better than one percent if this observable is to 
provide a test for charge symmetry violation in parton distributions.

\section*{Acknowledgements}

This work was supported in part [JTL] by National
Science Foundation research contract PHY0244822 and [AWT] by  
DOE contract DE-AC05-84150, under which SURA operates Jefferson 
Laboratory. 
  


\begin{thebibliography}{99}  

\bibitem{Miller} G. A. Miller, B. M. K. Nefkens and I. Slaus, Phys. Rep. 
       {\bf 194}, 1 (1990). 

\bibitem{Henley} E. M. Henley and G. A. Miller in {\it Mesons 
         in Nuclei}, eds M. Rho and D. H. Wilkinson 
         (North-Holland, Amsterdam 1979). 

\bibitem{Lon98} J. T. Londergan and A. W. Thomas, 
          Prog.\ Part.\ Nucl.\ Phys.\ {\bf 41}, 49 (1998). 

\bibitem{Lon04} J. T. Londergan and A. W. Thomas, 
          J. Phys.\ G Nucl.\ Part.\ Phys.\ {\bf 31}, 1 (2005). 

\bibitem{Londergan:2005xt}
  J.~T.~Londergan and A.~W.~Thomas,
  J.\ Phys.\ G {\bf 31}, 1151 (2005).
%
\bibitem{Mil05} G.A. Miller and A.W. Thomas, Int.~J. Mod.~Phys.~A{\bf 20}, 
   95 (2005). 

\bibitem{MRST03} A.D. Martin, R.G. Roberts, W.J. Stirling and 
	R.S. Thorne, Eur.~Phys.~J.~C{\bf 35}, 325 (2004). 

\bibitem{MRST01} A.D. Martin, R.G. Roberts, W.J. Stirling and 
	R.S. Thorne, Eur.\ Phys.\ J.\ C{\bf 23}, 73 (2002). 

\bibitem{Sat92} E. Sather, Phys. Lett.~B{\bf 274}, 433 (1992).

\bibitem{Rod94} E.N. Rodionov, A. W. Thomas and J. T. Londergan, 
          Mod.~Phys~Lett.\ A{\bf 9}, 1799 (1994). 

\bibitem{Londergan:2003ij}
  J.~T.~Londergan and A.~W.~Thomas,
  Phys.\ Rev.\ D {\bf 67}, 111901(R) (2003)
  [arXiv:hep-ph/0303155].

\bibitem{Londergan:2003pq}
  J.~T.~Londergan and A.~W.~Thomas,
  Phys.\ Lett.\ B {\bf 558}, 132 (2003)
  [arXiv:hep-ph/0301147].

\bibitem{MRST05} A.D. Martin, R.G. Roberts, W.J. Stirling and 
	R.S. Thorne, Eur.\ Phys.\ J.\ C{\bf 39}, 155 (2005). 

\bibitem{Glu05} M. Gluck, P. Jimenez-Delgado and E. Reya, 
	Phys.\ Rev.\ Lett.\ {\bf 95}, 022002 (2005). 

\bibitem{Dok77} Yu. L Dokshitzer, JETP {\bf 46}, 641 (1977). 

\bibitem{Gri72} V.N. Gribov and L.N. Lipatov, Sov.\ Journ.\ Nucl.\ 
   Phys.\ {\bf 15}, 78 (1972). 

\bibitem{Alt77} G. Altarelli and G. Parisi, Nucl.\ Phys.\ B{\bf 126}, 298 
   (1977). 

\bibitem{Bet57} H.A. Bethe and E.E. Salpeter, {\it Quantum Mechanics of 
One- and Two-Electron Atoms} (Academic Press, NY, 1957). 

\bibitem{GRV98} M. Gluck, E. Reya and A. Vogt, 
	Eur.~Phys.~J.\ C{\bf 5}, 461 (1998). 

\bibitem{Lon03} J. T. Londergan and A. W. Thomas, 
          Phys. Lett.~B{\bf 558}, 132 (2003).  

\bibitem{Lon05d} J. T. Londergan, D.P. Murdock and A. W. Thomas, 
       Phys. Rev. D{\bf72}, 036010 (2005); arXiv:hep-ph/0507029. 

\bibitem{Vig97} S.E. Vigdor, Second International Symposium on Symmetries 
   in Subatomic Physics, Seattle, WA, 1997 (unpublished).

\bibitem{Bor99} C. Boros, J. T. Londergan and A. W. Thomas,  
          Phys.~Rev.~D{\bf 59}, 074021 (1999). 
 
\bibitem{Ama91} NMC Collaboration, P. Amaudruz, M. Arneodo, A. Arvidson, 
 B. Badelek, G. Baum, J. Beaufays, I.G. Bird, M. Botje, C. Broggini, W. 
 Brueckner, A. Bruell, W.J. Burger, J. Ciborowski, R. Crittenden, R. van 
 Dantzig, H. Doebbeling, J. Domingo, J. Drinkard, A. Dzierba, H. Engelien, 
 M.I. Ferrero, L. Fluri, P. Grafstrom, D. von Harrach, M. van der Heijden, 
 C. Heusch, Q. Ingram, A. Jacholkowska, K. Janson, M. de Jong, E.M. Kabuss, 
 R. Kaiser, T.J. Ketel, F. Klein, B. Korzen, U. Kruener, S. Kullander, 
 U. Landgraf, F. Lettenstroem, T. Lindqvist, G.K. Mallot, C. Mariotti, G. van 
 Middelkoop, Y. Mizuno, J. Nassalski, D. Nowotny, N. Pavel, H. Peschel, C. 
 Peroni, B. Povh, R. Rieger, K. Rith, K. Roehrich, E. Rondio, L. Repolewski, 
 A. Sandacz, C. Scholz, R. Schumacher, U. Sennhauser, F. Sever, T.-A. Shibata, 
 M. Siebler, A. Simon, A. Staiano, G. Taylor, M. Treichel, J.L. Vuilleumier, 
 T. Walcher, K. Welch, R. Windmolders, and F. Zetsche, Phys.\ Rev.\ 
	Lett.\ {\bf 66}, 2712 (1991); Phys.\ Lett.\ B{\bf 295}, 159 
	(1992). 

\bibitem{Arn97} NMC Collaboration, M. Arneodo, A. Arvidson, 
 B. Badelek, M. Ballintijn, G. Baum, J. Beaufays, I.G. Bird, P. Bjoerkholm, 
 M. Botje, C. Broggini, W. Brueckner, A. Bruell, W.J. Burger, J. Ciborowski, 
 R. van Dantzig, A. Dyring, H. Engelien, M.I. Ferrero, L. Fluri, U. Gaul, 
 T. Granier, M. Grosse-Perdekamp, D. von Harrach, M. van der Heijden, 
 C. Heusch, Q. Ingram, M. de Jong, E.M. Kabuss, R. Kaiser, T.J. Ketel, 
 F. Klein, S. Kullander, U. Landgraf, T. Lindqvist, G.K. Mallot, C. Mariotti, 
 G. van Middelkoop, A. Milsztajn, Y. Mizuno, A. Most, A. Muecklich, J. 
 Nassalski, D. Nowotny, J. Oberski, A. Paic, C. Peroni, B. Povh, K. Prytz, 
 R. Rieger, K. Rith, K. Roehrich, E. Rondio, L. Repolewski, A. Sandacz, 
 D. Sanders, C. Scholz, R. Seitz, F. Sever, T.-A. Shibata, 
 M. Siebler, A. Simon, A. Staiano, M. Szleper, W. Tlaczala, Y. Tzamouranis, 
 M. Virchaux, J.L. Vuilleumier, T. Walcher, R. Windmolders, A. Witzmann, K. 
 Zaremba, and F. Zetsche, Nucl.\ Phys.\ 
	B{\bf 483}, 3 (1997). 

\bibitem{Sel97} CCFR Collaboration, W.G. Seligman, C.G. Arroyo, L. deBarbaro,
 P. deBarbaro, A.O. Bazarko, R.H. Bernstein, A. Bodek, T. Bolton, H. Budd,
 J. Conrad, D.A. Harris, R.A. Johnson, J.H. Kim, B.J. King, T. Kinnel, 
 M.J. Lamm, W.C. Lefmann, W. Marsh, K.S. McFarland, C. McNulty, S.R. Mishra,
 D. Naples, P.Z. Quintas, A. Romosan, W.K. Sakumoto, H. Schellman, F.J. 
 Sciulli, M.H. Shaevitz, W.H. Smith, P. Spentzouris, E.G. Stern, M. Vakili,
 U.K. Yang and J. Yu, Phys.~Rev.~Lett.~{\bf 79}, 1213 (1997). 

\bibitem{Yan01} CCFR Collaboration, U.K. Yang, T. Adams, A. Alton, C.G. 
 Arroyo, L. deBarbaro,P. deBarbaro, A.O. Bazarko, R.H. Bernstein, A. Bodek, 
 T. Bolton, J. Brau, D. Buchholz, H. Budd, L. Bugel, J. Conrad, R.B. Drucker, 
 B.T. Fleming, J.A. Formaggio, R. Frey, J. Goldman, M. Goncharov, D.A. Harris, 
R.A. Johnson, J.H. Kim, B.J. King, T. Kinnel, S. Koutsoliotas,  
 M.J. Lamm, W. Marsh, D. Mason, K.S. McFarland, C. McNulty, S.R. Mishra,
 D. Naples, P. Nienaber, A. Romosan, W.K. Sakumoto, H. Schellman, F.J. 
 Sciulli, M.H. Shaevitz, W.H. Smith, P. Spentzouris, E.G. Stern, 
 N. Suwonjandee, A. Vaitaitis, M. Vakili, J. Yu, G.P. Zeller and E.D. 
 Zimmerman, Phys.~Rev.~Lett.~{\bf 86}, 2742 (2001). 

\bibitem{Boros:1999fy}
  C.~Boros, F.~M.~Steffens, J.~T.~Londergan and A.~W.~Thomas,
  Phys.\ Lett.\ B {\bf 468}, 161 (1999)
  [arXiv:hep-ph/9908280].

\bibitem{Bar76} R.M. Barnett, Phys.~Rev.~Lett.~{\bf 36}, 1163 (1976).

\bibitem{Geo76} H. Georgi and H.D. Politzer, Phys.~Rev.~D{\bf 14}, 1829 (1976).

\bibitem{DurWeb} The Durham parton distributions are obtained at  
	http://www-spires.dur.ac.uk/hepdata/mrs.html  

\bibitem{MRST04} A.D. Martin, R.G. Roberts, W.J. Stirling and 
	R.S. Thorne, Phys.~Lett.~B{\bf 604}, 61 (2004). 

\end{thebibliography}
\end{document}